\documentclass[aps,prl,letterpaper,10pt,twocolumn,showpacs,superscriptaddress]{revtex4-1}
\usepackage[utf8]{inputenc}
\usepackage{hyperref}
\usepackage{amsmath,revsymb,amssymb}
\usepackage{graphicx}
\usepackage[colorinlistoftodos]{todonotes}
\usepackage{color}
\usepackage{bbold}
\usepackage{braket,ulem}

\newcommand{\tdown}{t_{\rm down}} 
\newcommand{\trx}{t_{\rm rx}}
\newcommand{\tdiff}{t_{\rm diff}}

\begin{document}

\title{
Towards Quantum Communication from Global Navigation Satellite System}

\author{Luca Calderaro}
\affiliation{
Dipartimento di Ingegneria dell'Informazione, Universit\`a di Padova, via Gradenigo 6B, 35131 Padova, Italy.}
\affiliation{
 Istituto Nazionale di Fisica Nucleare (INFN) - sezione di Padova, Italy}
\author{Costatino Agnesi}
\affiliation{
Dipartimento di Ingegneria dell'Informazione, Universit\`a di Padova, via Gradenigo 6B, 35131 Padova, Italy.}
\affiliation{
 Istituto Nazionale di Fisica Nucleare (INFN) - sezione di Padova, Italy}
\author{Daniele Dequal}
\affiliation{Matera Laser Ranging Observatory, Agenzia Spaziale Italiana, Matera, Italy}
\author{Francesco Vedovato}
\affiliation{
Dipartimento di Ingegneria dell'Informazione, Universit\`a di Padova, via Gradenigo 6B, 35131 Padova, Italy.}
\affiliation{
 Istituto Nazionale di Fisica Nucleare (INFN) - sezione di Padova, Italy}
\author{Matteo Schiavon}
\affiliation{
Dipartimento di Ingegneria dell'Informazione, Universit\`a di Padova, via Gradenigo 6B, 35131 Padova, Italy.}
\affiliation{
 Istituto Nazionale di Fisica Nucleare (INFN) - sezione di Padova, Italy}
 \author{Alberto Santamato}
\affiliation{
Dipartimento di Ingegneria dell'Informazione, Universit\`a di Padova, via Gradenigo 6B, 35131 Padova, Italy.}
\author{Vincenza Luceri}
\affiliation{e-GEOS spa, Matera, Italy}
\author{Giuseppe Bianco}
\affiliation{Matera Laser Ranging Observatory, Agenzia Spaziale Italiana, Matera, Italy}
\author{Giuseppe Vallone}
\affiliation{
Dipartimento di Ingegneria dell'Informazione, Universit\`a di Padova, via Gradenigo 6B, 35131 Padova, Italy.}
\affiliation{
 Istituto Nazionale di Fisica Nucleare (INFN) - sezione di Padova, Italy}
\author{Paolo Villoresi}
\affiliation{
Dipartimento di Ingegneria dell'Informazione, Universit\`a di Padova, via Gradenigo 6B, 35131 Padova, Italy.}
\affiliation{
 Istituto Nazionale di Fisica Nucleare (INFN) - sezione di Padova, Italy}



\date{\today}

\begin{abstract}
Satellite-based quantum communication is an invaluable resource for the realization of a quantum network at the global scale. In this regard, the use of satellites well beyond the low Earth orbits gives the advantage of long communication time with a ground station. However, high-orbit satellites pose a great technological challenge due to the high diffraction losses of the optical channel, and the experimental investigation of such quantum channels is still lacking. Here, we report on the first experimental exchange of single photons from Global Navigation Satellite System at a slant distance of 20000 kilometers, by exploiting the retroreflector array mounted on GLONASS satellites. We also observed the predicted temporal spread of the reflected pulses due to the geometrical shape of array. Finally, we estimated the requirements needed for an active source on a satellite, aiming towards
quantum communication from GNSS with state-of-the-art technology.
\end{abstract}

\maketitle

\section{Introduction}
Satellite-based technologies are the enabling tools for a wide range of civil, military and scientific applications \citep{SpringerApplications,Roddy2006,SAGE2009,Tan2014}, like communications, navigation and timing, remote sensing, meteorology, reconnaissance, search and rescue, space exploration and astronomy.  In particular, Global Navigation Satellite Systems (GNSS)  were developed in the second half of XX century to provide autonomous geo-localization by exploiting a network of satellite exchanging position- and time-information with different locations on Earth \citep{SpringerGNSS}. The strategic importance of such infrastructure led different countries to deploy their own GNSS constellations, e.g. the American Global Positioning System (GPS), the Russian GLONASS, the European Galileo, the Chinese BeiDou, the Japanese QZSS and the Indian INRSS/NAVIC. The very core of these navigation systems is the capability of safely transmitting information and data from orbiting satellites to several ground stations on Earth by exploiting radio \citep{Roddy2006} or optical communications \citep{Hemmati2009}. In fact, the protection of such infrastructure from a malicious adversary is of crucial importance for both civil and military operations, representing a critical issue that is continuously and extensively under development.

\begin{figure*}[t] 
\includegraphics[width=1\textwidth,clip]{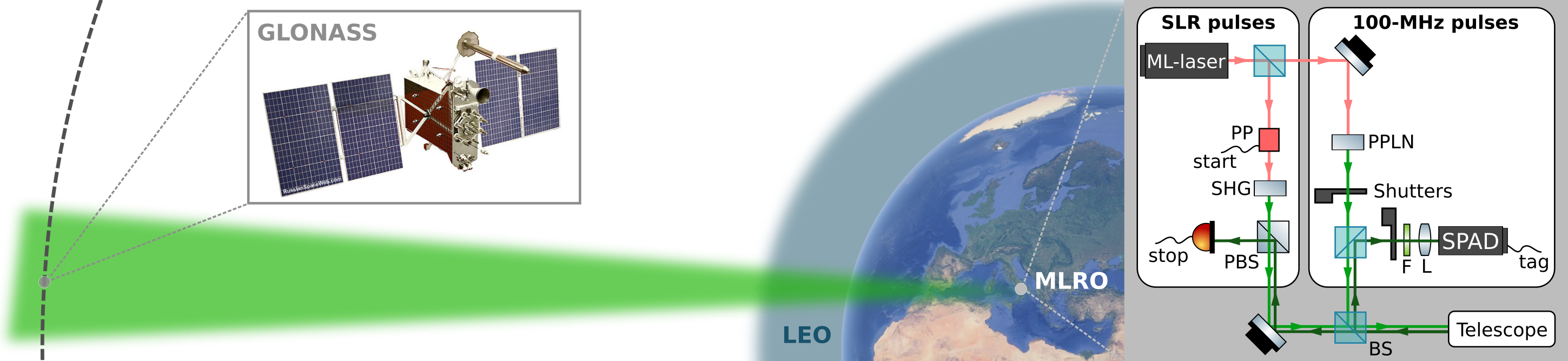}   
\caption{Illustrative representation (to scale) of the typical distance ($\approx 20000$~km) between GLONASS terminals and the MLRO ground station in Italy. GNSS constellations are well above the LEO satellites (maximum altitude about 2000 km). In the right panel it is shown the setup used to experimentally study the optical channel (see the main text). In particular, the communication protocol starts with the SLR start signal at $t=0$~ms. The 100-MHz pulses are sent to the satellite from $t=0$~ms to $t=100$~ms, opening the shutter placed in the transmission path. At $t=100$~ms a second SLR pulse is sent to the satellite and after $5$~ms the receiving shutter open the receiving path till $t=180$~ms. Image of Glonass-K satellite taken from \href{http://www.russianspaceweb.com/uragan_k.html}{Russian SpaceWeb}.}
    \label{fig:illustration}
\end{figure*}

At the same time, space quantum communications (QCs) (see the reviews \citep{Bedington17, Khan2018}) represent a promising resource to guarantee unconditional security for satellite-to-ground \citep{Rarity02,Villoresi2004} and inter-satellite optical links \citep{Pfennigbauer2003,Tomaello2011}, by exploiting quantum information protocols as quantum key distribution (QKD) \citep{Gisin02,Scarani09}.  Despite being a relative new research field born in the 2000s, satellite QCs developed rapidly from the first experimental studies of photon-exchange \citep{Villoresi08,Yin2013} from Low-Earth-Orbit (LEO) satellites to the feasibility-test of different photon encodings such polarization \citep{Vallone15,Takenaka17}, time-bin \citep{Vallone16}, and continuous-variable \citep{Gunthner17}. Then, the strong efforts of the Chinese Academy of Sciences led to full in-orbit demonstrations of such technologies last year \citep{Yin17-1,Ren,Liao17,Yin17-2}, culminated in the realization of an intercontinental quantum-secure communication between China and Austria \citep{Liao18}.

Due to optical losses, most of the demonstrations of satellite QCs were limited, so far, to LEO satellites.
However, the high orbital velocity of LEO satellites limits their visibility periods from the ground station, and subsequently the time available for QCs to just few minutes per passage.
Conversely, the use of satellites at higher orbits can greatly extend the communication time, reaching few hours in the case of GNSS. Furthermore, QCs could offer interesting solutions for GNSS security for both satellite-to-ground and inter-satellite links, offering novel and unconditionally secure protocols for the authentication, integrity and confidentiality of exchanged signals. For example, a GNSS inter-satellite network for QC has already been proposed to strengthen the security of the Galileo architecture \citep{Gerlin13}.  
This would allow the generation of cryptographic keys and the construction of a secure satellite QKD-network, thus preventing the catastrophic consequences of malicious hijacking of GNSS satellites.

Here, we experimentally demonstrate the feasibility of QC  between a GNSS terminal and a ground station, over a channel length of about 20000 km by using current technology. We report on the first exchange of few photons per pulse  between two different satellites of GLONASS constellation and the Space Geodesy Centre of the Italian Space Agency in Matera,  Italy, by exploiting the passive retro-reflectors mounted on the satellites. By estimating the actual losses of such a channel, we can evaluate the characteristics of both a dedicated quantum payload and a receiving ground station, hence attesting the feasibility of QC from GNSS in terms of achievable signal-to-noise ratio and detection rate. Our work extends the limit of long-distance free-space single-photon exchange, which  was demonstrated so far with a channel length of about 7000 km by exploiting a Medium-Earth-Orbit (MEO) satellite \citep{Dequal16}.

\section{The optical setup}
The feasibility of QC from GNSS orbits was studied  experimentally using some of such satellite equipped with an array of corner-cube retroreflectos (CCRs). A weak source in  GNSS orbit is emulated by exploiting the two-way scheme already tested with LEO \cite{Vallone15} and MEO satellites \cite{Dequal16}. Our scheme takes advantages of Satellite Laser Ranging (SLR) technique, in which bright laser pulses sent to CCRs are used to accurately measure the distance of such satellites for geodynamics purposes \citep{Degnan93}. 

The experiment presented here was performed at the 
Italian Space Agency's
Matera Laser Ranging Observatory (MLRO)  by using the setup sketched in Figure~\ref{fig:illustration}. The observatory is a SLR station equipped with a mode-locking Nd:YVO$_4$ laser oscillator (ML-laser), operating at 1064 nm  with 100 MHz repetition rate and paced by an atomic clock. The SLR pulses (wavelenght, 532 nm; energy, $\sim$100 mJ; repetition rate, 10 Hz) are obtained by selecting one seed pulse every $10^7$ with a pulse-picker (PP), which is then amplified and up-converted via a second-harmonic-generation (SHG) stage. The SLR pulses are sent to the targeted satellites equipped with CCRs  \citep{Degnan93} by using the 1.5-m diffraction-limited Cassegrain telescope of MLRO \citep{Bianco01}. Then, after the reflection by the orbiting terminals the pulses are collected by a fast analog micro-channel plate detector (Hamamatsu R5916U-50) placed after a 50:50 polarizing beam-splitter (PBS) used to separate the transmitted beam from the received one. A dedicated time-tagger with sub-picosecond accuracy recorded the start and stop signals generated by the PP and the detector respectively. The single-shot measurement of the satellite distance is then estimated from the time-difference of these two signals, i.e. the  \textit{round-trip-time},  with an error below 20 ps. 

A setup dedicated to the feasibility study of QC from GNSS was implemented in parallel to the SLR system. The same laser oscillator is used to produce a 100-MHz pulse-train with wavelength $\lambda = 532$ nm, $\sim$1 nJ of energy and 55 ps of pulse duration at full-width-half-maximum (FWHM) by exploiting a 50 mm long periodically poled lithium niobate (PPLN) non linear crystal from HC Photonics. This beam, synchronized with the SLR pulse-train, is combined with the outgoing SLR pulses by using a 50:50 beam splitter (BS) and the two light beams are sent to the targeted GNSS satellites.
The receiving apparatus of the 100-MHz beam is comprised of a 50:50 BS to separate the outgoing and ingoing beam, a $3$~nm FWHM spectral filter (F) with transmission band centered at $532$~nm, a focusing lens (L) and a silicon single photon avalanche detector (SPAD), provided by Micro-Photon-Devices (MPD, Italy), with  $\sim$50$\%$ quantum efficiency, $\sim$400 Hz dark count rate and $40$~ps of jitter. The time  of arrival of the returning photons (tag) is recorded with $1$~ps resolution by the  time-to-digital converter QuTAG from QUTOOLS. 

We implemented a communication protocol to separate the transmitting and receiving phases by using two mechanical shutters. Since the round trip time of photons reflected by GNSS satellites is around $130$ ms, the total period of the communication protocol is 200 ms.
In the first half, the transmitting (receiving) shutter is open (close) and the 100-MHz pulses are transmitted. Vice-versa, in the second half the the receiving (transmitting) shutter is open (close) and the 100-MHz pulses coming from the satellite can be detected. 

Due to optical diffraction, atmospheric absorption, and the finite sizes of the cross-section and active area of the CCR-array, the beams are attenuated by several orders of magnitude after being retro-reflected. As a result the reflected pulses of the 100-MHz train have a low mean number of photons $\mu_{\mathrm{sat}}$ at the satellite reflection, thus emulating an active weak source placed on a GNSS terminal orbiting 20000 km away from the MLRO ground station. In the following sections a detailed estimation of the channels losses and of $\mu_{\mathrm{sat}}$ is provided by measuring the  detection rate of the returning photons, attesting the faithfulness of our implementation. The targeted GNSS satellites are two different terminals of the GLONASS constellation, namely Glonass-134 and Glonass-131 (Space Vehicle Number: 802 and 747, respectively).

\section{Model of the channel losses}
In our experiment, the mean photon number per pulse $\mu_\mathrm{sat}$ emitted by the simulated source at the satellite is not known \textit{a priori}.  We can estimate it \textit{a posteriori} as  $\mu_{\mathrm{sat}} = R_{\mathrm{det}} / (\nu_{\mathrm{tx}} \tdown \trx)$, by experimentally evaluating the detection rate $R_{\mathrm{det}}$, and by knowing the down-link transmittance $\tdown$, the transmittance  of the receiving apparatus $\trx$ and the repetition rate $\nu_{\rm tx}=100$~MHz of the source. Losses are expressed in dB as $l=-10\log_{10}t$, where $t$ is the transmittance. The receiver losses are promptly estimated taking into account the reflection and transmission losses through all the optical elements (8.8 dB) and the  detector quantum efficiency (3 dB).
 
The down-link channel losses can be evaluated as the product of the atmospheric transmission $t_{\mathrm{a}}$ and the geometrical transmission due to diffraction $\tdiff$. We follow two independent approaches for estimating the transmission due to diffraction and compares the results for the validation of the model.  The targeted GNSS satellites are part of different generations, GLONASS-K1 for Glonass-134 and GLONASS-M for Glonass-131, both equipped with a planar array of CCRs, with circular and rectangular shape respectively~\citep{Govus17}. Their CCRs are characterized by the absence of  coating on the reflecting faces, such that the light is back reflected by total internal reflection (TIR). This implies a far field diffraction pattern (FFDP) which is quite different from the simple Airy disk given by a circular aperture~\citep{Murphy13}. The FFDP of a TIR corner cube has a central Airy-like disk, with $26.4\%$ reduced central intensity peak from the circular aperture with equivalent area, surrounded by six lobes placed on the vertices of a hexagon. The lobes are displaced from the center of the FFDP by $\theta_d \approx 1.4 \lambda/D_\mathrm{CCR}$, with $D_\mathrm{CCR}=26$~mm the CCR diameter \citep{GlonassK1}, corresponding to a displacement  $\theta_d \approx 29$~$\mu$rad. Since the velocity aberration of GNSS satellite is around $26$~$\mu$rad~\citep{Degnan93}, the MLRO telescope is receiving the lateral lobes of the FFDP.
 In particular, the lateral lobes have an intensity which is  $\approx 30\%$ of the central peak.  Since the central intensity peak $I_0$ of a circular aperture of area $A$ depends on the power $P_0$ incident on it via $I_0 = P_0 A/(\lambda^2 R^2)$, with $R$ the distance from the aperture, the transmission due to diffraction can be evaluated by
\begin{equation} \label{eq:murphy}
\tdiff = 0.264\cdot 0.3 \frac{A_\mathrm{CCR} A_\mathrm{tel}}{\lambda^2 R^2},
\end{equation}
where $A_{CCR}$ and $A_\mathrm{tel}$ are the areas of the CCR and the ground telescope, respectively ~\citep{Murphy13}.

An alternative approach is given in~\citep{Dequal16} in which the FFDP is approximated as a top-hat pattern with solid angle $\Omega$, so that the diffraction transmittance is evaluated as $A_\mathrm{tel} / (\Omega R^2)$. Since the solid angle can be estimated by the array cross-section $\Sigma$~\citep{Degnan93,Arnold79}, we have that
\begin{equation} \label{eq:degnan}
\tdiff  = \frac{\Sigma}{ 4\pi   \rho A_\mathrm{RRA}} \frac{A_\mathrm{tel}}{R^2},
\end{equation}
being $\rho = 0.93$ the reflectivity of the uncoated CCR and $A_\mathrm{RRA}$ the array effective area.

In clear sky conditions, the losses due to atmospheric transmission for the used $\lambda$ is $l_{\mathrm{a}} \approx  0.4$ dB \citep{Degnan93} and considering a satellite slant distance $R \approx 20000$ km, the predicted down-link channel losses are $l_{\rm down}\approx 62$ dB, from  both models Eqs.~\ref{eq:murphy} and \ref{eq:degnan} to estimate the diffraction losses. This assessment of the channel losses allow us to experimentally estimate $\mu_{\mathrm{sat}}$ by measuring the detection rate, as presented in the following section.

\section{Detection of single photons from GNSS terminals}
In this section, we present the results obtained with the two  satellites above described. 
We provide here a detailed analysis of the data obtained from Glonass-134. The case of Glonass-131,
in which we used also a single-photon photomultiplier (PMT)  in parallel to the SPAD to compare their performances, is described in the Supplementary Material (SM). In a single passage of Glonass-134, we had two distinct acquisitions separated by almost one hour corresponding to the maximum and minimum distance of the satellite from MLRO. In particular, the first acquisition lasted about 2 minutes, with mean slant distance of about 20200 km, whereas the second one lasted about 5 minutes, with mean slant distance of 19500 km. 

\begin{figure}[t]
	\includegraphics[width=\columnwidth,clip]{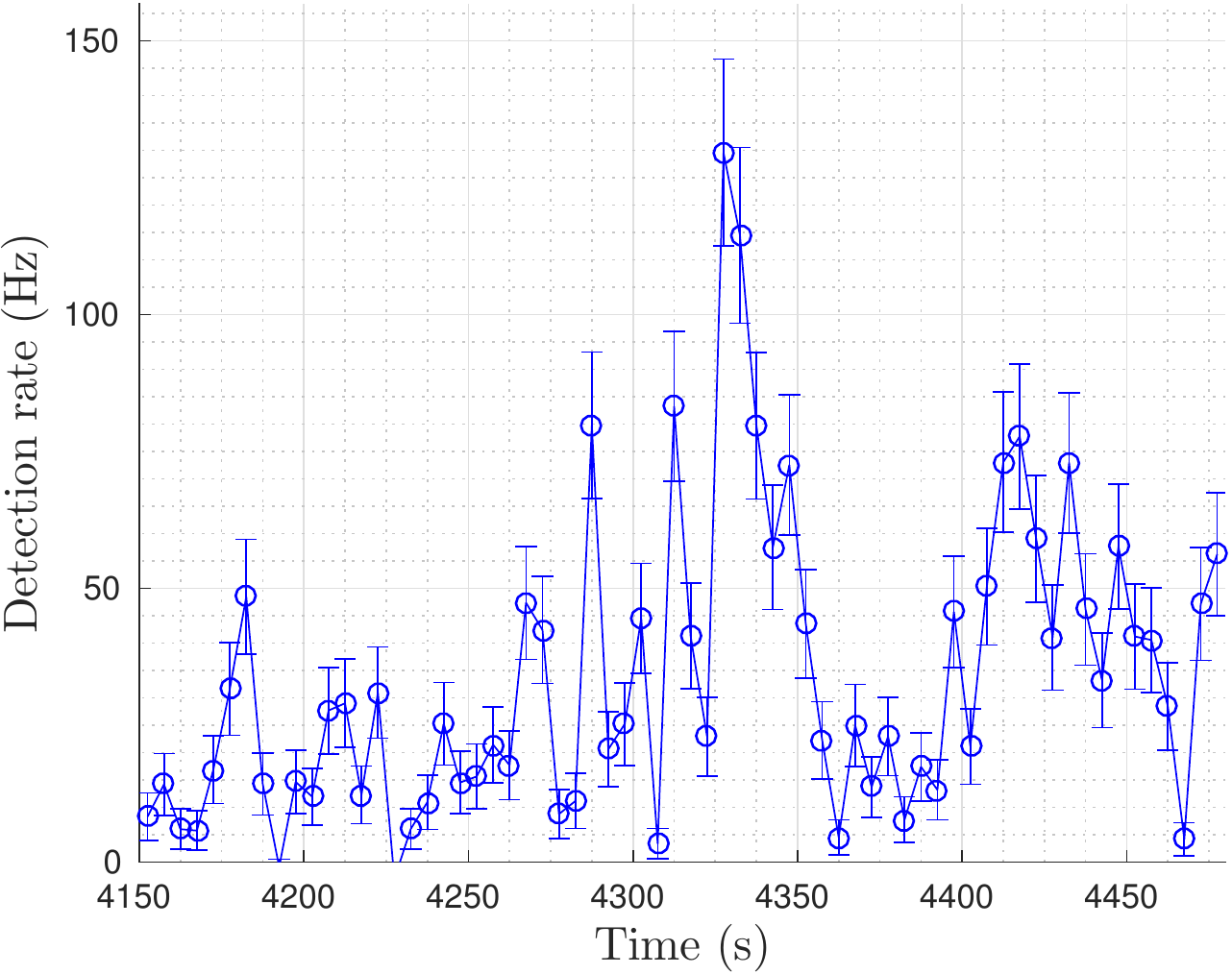}
    \caption{Detection rate from Glonass-134 at 19500 km slant distance. Each point is calculated integrating over an acquisition time window $\mathcal I_k$ of $\tau = 5$~s.}
    \label{fig:G134-2_det_rate}
\end{figure}

\begin{figure}[t]
	\includegraphics[width=\columnwidth,clip]{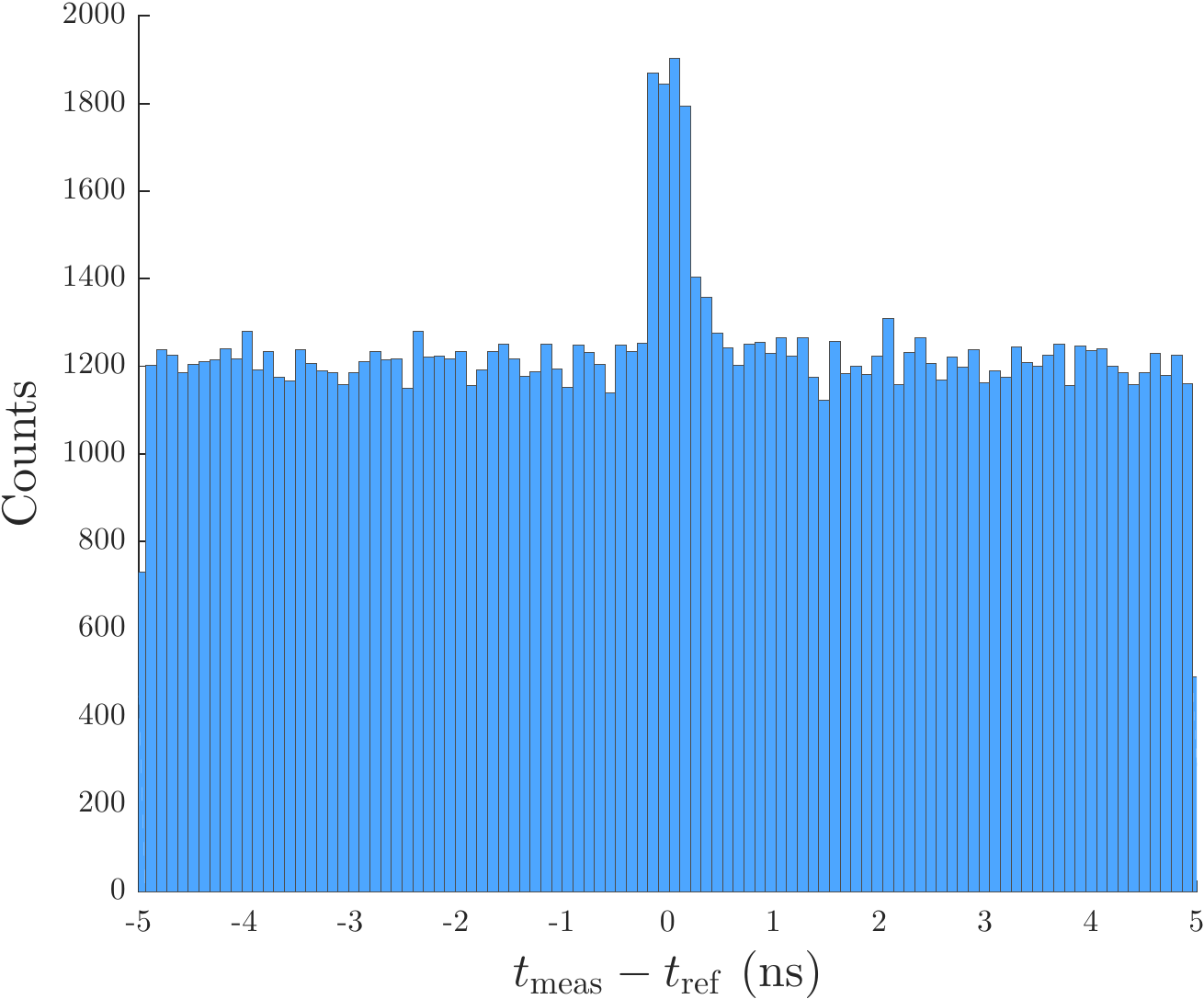}
    \caption{Histogram of residuals between the measured and the expected time of arrival of the photons, from Glonass-134 at a slant distance of 19500 km. Here, we consider acquisition time windows $\mathcal I_k$ with detection rate $R_s > 30$~Hz. Each bin is $100$~ps wide.}
    \label{fig:G134-2_hist}
\end{figure}

In Fig.~\ref{fig:G134-2_det_rate} we show the signal detection rate from Glonass-134 for the second acquisition (the results for the first acquisition are analogue). The detection rate was estimated in the following way. We divided the whole acquisition in time intervals ${\mathcal I}_k$ of duration $\tau = 5$~s. For each interval we made the histogram (see Figure~\ref{fig:G134-2_hist}) of the time difference between the tagged detection $t_{\mathrm{meas}}$ and the expected time of arrival of the photon $t_\mathrm{ref}$,  estimated from  the SLR acquisition performed in parallel \citep{Vallone15,Dequal16}.  Then, we chose a time window $ w = 400$~ps, centered around $t_\mathrm{ref}$, and estimated the number of  photon detections $N_\mathrm{det}$ as the difference of the total and background counts  within $w$, which was chosen much larger than the detector jitter  ($\approx  40$~ps) since the retroreflected pulses are temporally spreaded by the CCR array. The background was uniformly distributed within the 10 ns period between two sent pulses {(see Figure~\ref{fig:G134-2_hist})}, therefore we estimated its rate counting the detections over a time window which is at least $1$~ns away from $t_\mathrm{ref}$. Finally, the signal detection rate was obtained via $R_\mathrm{det} = N_\mathrm{det} / (\tau \delta)$ where $\delta=0.3$ is the duty cycle of the communication protocol.
Then, we discarded the time windows ${\mathcal I}_k$ with $R_\mathrm{det} < 30$~Hz, to filter out acquisition with  low  signal-to-noise ratio (SNR). 
Such selected time windows gave the integrated histogram shown in Fig.~\ref{fig:G134-2_hist}. 

At the end of such analysis, we obtained a mean detection frequency $\overline{R}_\mathrm{det} \approx 58$~Hz, a SNR of $0.53$ and mean number of photons at the satellite $\overline{\mu}_{\mathrm{sat}} \approx 14.5$ for the second acquisition of Glonass-134.
In the same way we analyzed the first acquisition of the same passage, obtaining a mean detection frequency $\overline{R}_\mathrm{det} \approx 59$~Hz, a SNR of $0.41$ and a mean number of photons at the satellite $\overline{\mu}_{\mathrm{sat}} \approx 16.1$. In this case we used a signal time window $w$ of $600$~ps due to the larger temporal spread. The results are summarized in Tab.~\ref{tab:results}, along with the acquisition of Glonass-131 detailed in the SM.

\begin{figure}
	\includegraphics[width=\columnwidth,clip]{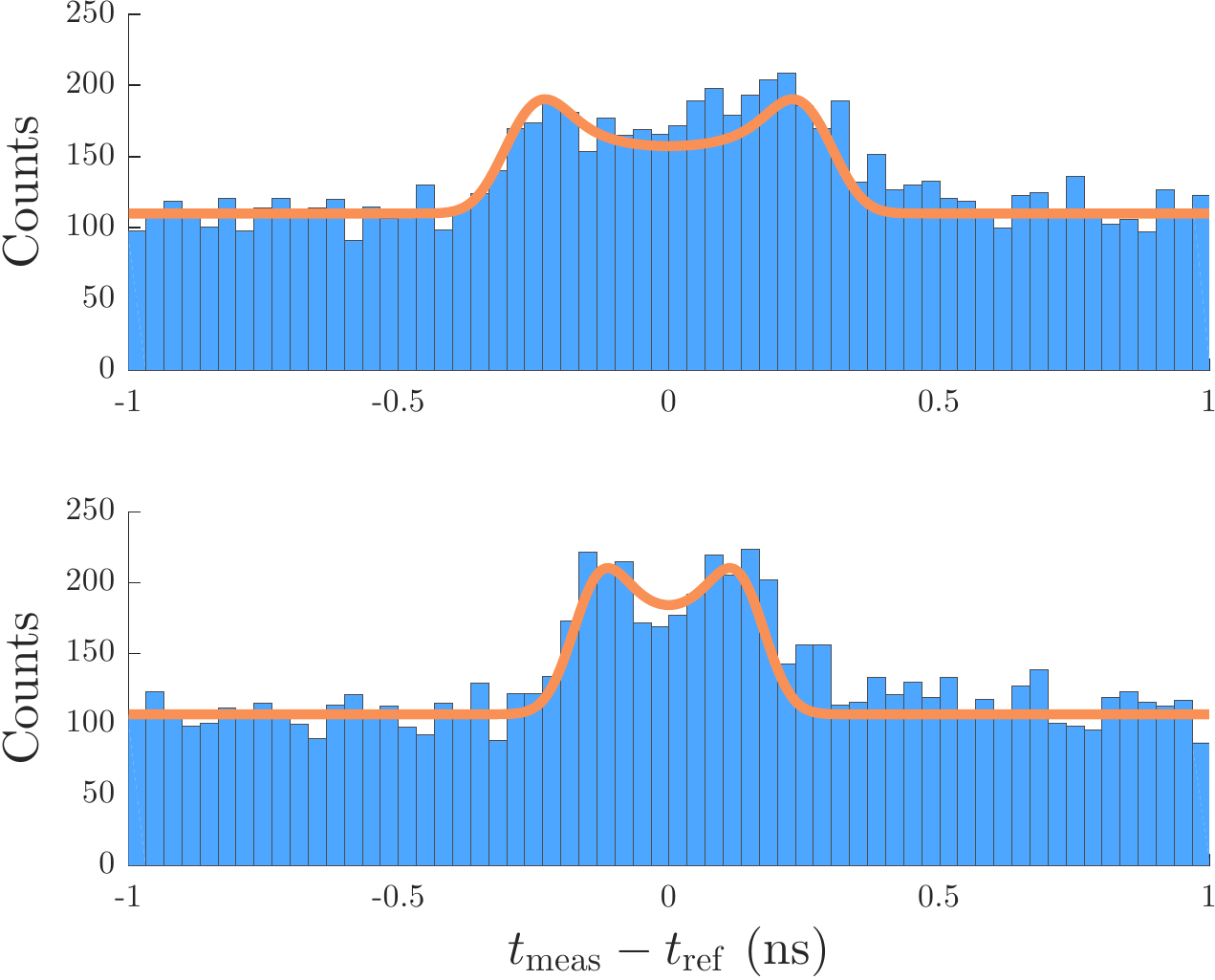}
    \caption{Histogram of residuals between measured and expected time of arrival of the photons for Glonass-134 at 20200 km (\textbf{top}) and 19500 km (\textbf{bottom}). The incident angle of the beam on the array is about $9 \deg$ and $5 \deg$, respectively. Here we integrate on time windows $\mathcal I_k$ with high detection rate to highlight the temporal spread of the back reflected impulse. Based on our model, the temporal peak to peak distance is about $430$~ps and $250$~ps for incident angles of $9 \deg$ and $5 \deg$, respectively (continuous lines).}
    \label{fig:doublepeak}
\end{figure}

It is worth noticing that, with our apparatus, it is possible to resolve the temporal distribution of the returning pulse given by the particular design of the CCR array,  hence revealing the ``signature'' of Glonass-134 that is equipped with a holed circular CCR array~\citep{GlonassK1}. If the incident angle on the array is not zero, the pulses reflected by the CCRs closer to the ground station have a smaller round trip time with respect to the further CCRs, resulting in a temporal spread of the pulse. We simulated the temporal shape of the pulse for incident angles $5 \deg$ and $9 \deg$, corresponding to the incident angles of the two acquisitions, and compare them with the actual data in Fig.~\ref{fig:doublepeak}. From the simulation, the corresponding temporal peak-to-peak distance is $250$ ps and $430$ ps  for the two acquisitions respectively in agreement with the experimental estimation. The simulation is performed supposing that the single CCR does not change the temporal shape of the pulse but introduces a temporal offset depending on its position in the array and on the incident angle of the beam. Using a pulse with $100$ ps of FWHM and summing up the contributions of all CCRs we obtained the shapes depicted in the figure.
It is worth noticing that the continuous lines shown in Fig.~\ref{fig:doublepeak} are obtained by such a-priori model (adding the measured background) and not by fitting the data.

As shown in the work by Otsubo \textit{et al.} \citep{Otsubo01}, GLONASS flat CCRs array exhibits particular temporal distribution determining higher error in the laser ranging measurement, in which the mean number of photons at the receiver is usually much greater than one. The authors of \citep{Otsubo01} observed the ``signature'' of the GLONASS satellites by integrating one year of data acquisition.  On the contrary, our  result shows that using single photons detectors and high repetition source the temporal distribution of the pulse can be measured, even with low mean number of photon at the satellite and short data integration time.
A more accurate measurement could be done using a mean number of photons at the receiver about one, but this is beyond the scope of this work. We note that this measurement could even be used to increase the accuracy in the determination of the orientation of the array and hence the attitude of the satellite, which is of critical importance for the processing of GNSS data \cite{Pear2002,Mont2015}.

\begin{table*}
\centering
\begin{tabular}{c|c|c|c|c|c|c|c}
  Satellite passage & Slant distance (km) & Detector & $\overline{R}_{\mathrm{det}}$ (Hz)& SNR & $\overline{\mu}_{\mathrm{sat}}$  & $l_{\mathrm{down}}$ (dB) & $l_{\mathrm{rec}}$ (dB) \\ 
 \hline\hline
 Glonass-134 & 19500 & SPAD & 58 & 0.53 & 15 & 62.1  & 11.8  \\
  & 20200 & SPAD & 59 & 0.41 & 16 & 62.5  & 11.8  \\ \hline
 Glonass-131 & 20250 & SPAD & 27  & 0.43 & 15 & 62.6  & 14.8  \\
 & & PMT & 6  & 0.21 & 16 & 62.6  & 21.8 \\
\end{tabular}
\caption{Summary of the results. Mean signal detection rate $\overline{R}_{\mathrm{det}}$,  mean photon number at the satellite $\overline{\mu}_{\mathrm{sat}}$, mean down-link  losses $l_{\mathrm{down}}$.}
\label{tab:results}
\end{table*}

\section{Towards quantum communication from GNSS}
Based on these results, we can estimate the performance of a transmitter and receiver needed for the realization of quantum communication from GNSS satellites. For practical quantum communication we target a SNR larger then $100$ and a detection rate larger than $10$ kHz.

\begin{figure}
	\includegraphics[width=\columnwidth,clip]{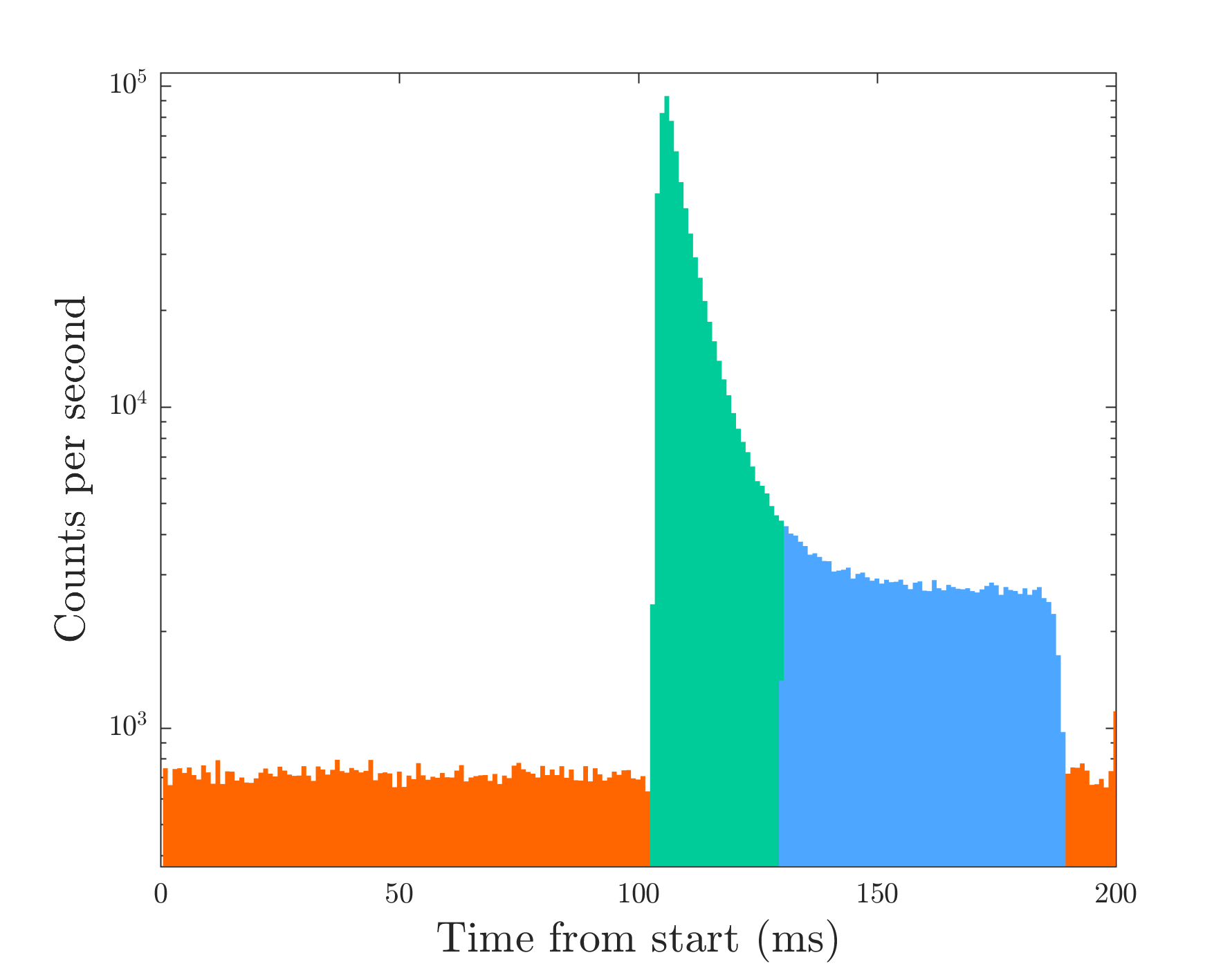}
    \caption{Stacked histogram of detection rate with respect to start signal for the second acquisition of Glonass-134. This histogram shows the duty cycle of the communication protocol. Orange bars show the detection rate when the receiving shutter is closed. Blue and Green bars shows detections when the receiving shutter is open. In particular the Blue bars are detections that occurs a RTT after the start signal. }
    \label{fig:G-134_200ms}
\end{figure}

At the receiver, the background affecting the SNR can be significantly reduced with respect to present experiment in a dedicated QC application. In our work, the background was estimated by using the detections distribution on the $200$~ms period, which is shown in Fig~\ref{fig:G-134_200ms} for the passage of Glonass-134 at $19500$~km. The blue bars corresponds to the counts in which we expect the transmitted photons to arrive at the detector. This time window starts one RTT from the first transmitted qubit and ends when the receiving shutter is closed. A large part of these counts are due to noise. The intrinsic dark count rate of the detector amounts to $N_{\mathrm{rx}} = 700$ Hz. They are estimated in the first $100$~ms of the period, when the receiving shutter is closed. This noise could be almost halved, reaching the intrinsic dark count rate of the detector, by optimizing the optical isolation of the detector from the room light.
Another source of noise is the fluorescence that occurs when the upgoing SLR pulse passes through the optical elements in common with our optical path. The intensity of the fluorescence light reduces exponentially in time with half-life that depends on the material. A remaining tail is included in the blue region and amounts to $N_{\mathrm{fluo}} = 195$ Hz. This noise can be eliminated, since this pulse is useless for the protocol and can just be avoided.
The remaining, and predominant, detections are due to satellite albedo and background of the field of view. This noise is uniformly distributed in time in the blue region and amounts at $N_{\mathrm{alb}} = 1.9$~kHz. We can reduce it by an order of magnitude using a bandpass filter of $0.3$~nm instead of $3$~nm. 
Moreover, a dedicated receiver may avoid the use of signal losses due to beam splitters. Indeed the satellite tracking may be done using a different wavelength. With respect to our setup, this would enhance four times the signal, although correspondingly augmenting $N_{\mathrm{alb}}$. Adopting these solutions at the receiver we expect a SNR and a detection rate raised of a factor 10 and 4, respectively.

Regarding the transmitter, we consider an active source on the satellite with a mean photon number per pulse close to 1. Compared to the current result, this involves a signal reduction of about a factor $15$. However, the down-link coupling efficiency can be greatly enhanced by using an appropriate telescope. We consider a down-going beam with 10~$\mu$rad of semi angular aperture, shrinking the beam spot on ground and using a point ahead to compensate for velocity aberration
as recently demonstrated in \cite{Liao17}. This would reduce the diffraction losses of $20$ dB with respect to the channel losses estimated above. The temporal spread due to the reflector array would not be present, allowing for a narrower temporal filter $w$ that could be chosen considering only the jitter of the detector ($\approx 40$~ps). Moreover, with $40$~ps jitter, the repetition rate could be increased  to  more than $1$~GHz, thus enhancing the detection rate.  With these expedients, the expected SNR and  detection rate are of the order of 100 and 10 kHz, respectively.

\section{Conclusions} 

This work demonstrates the first exchange of few photons per pulse  ($\mu_{sat} \simeq 10$) along a channel length of 20000 km, from Glonass-134 and Glonass-131 to MLRO, reaching a SNR  about 0.5 and a detection rate  around 60~Hz. We evaluated the requirements needed for a transmitter mounted on a GNSS satellite and a ground receiver for the realization of QC between the two terminal. Our findings demonstrate that QC from GNSS satellite is feasible with current state-of-the-art technology.

 Extending QC to GNSS is of primary importance for secure communications at the global scale, as discussed above,  but it is also 
a resource for fundamental tests of physics in space. Indeed, QC from satellite opens the possibility of testing the foundations of quantum mechanics in the space scenario, as envisaged in theoretical studies \citep{Rideout12} and mission proposals \citep{Ursin09,Scheidl13,Jennewein14}, and already realized in actual implementations \citep{Yin17-1,Ren,Vedovato17} at the LEO distance. A channel length of over 20000 km could enable the design of new experiments that test the validity of quantum mechanics at higher orbits and permit the use of satellites following highly elliptical orbits. Such orbital characteristics might be of key importance to observe  gravity-induced effects on quantum interference, \citep{Brodutch2015,Bruschi14,Zych2012}, that could shed light on  the interplay between general relativity and quantum mechanics, thus validating physical theories and placing bounds on phenomenological models.
Concluding, our results  pave the way for new applications of quantum technologies and fundamental experiments of physics exploiting QC from high-orbit satellites,
which may be implemented on next-generation GNSS constellation.

\begin{acknowledgments}
We would like to thank Francesco
Schiavone, Giuseppe Nicoletti, and the MRLO technical
operators for the collaboration and support.
We acknowledge the International Laser Ranging Service (ILRS) for the satellite data. 
L.C. and F.V. acknowledge the Center of Studies and Activities for Space (CISAS) ``Giuseppe Colombo'' for
financial support. Our research was partially funded by the Moonlight-2 project of INFN.
\end{acknowledgments}

\section{Supplementary Material}

\begin{figure}
	\includegraphics[width=\columnwidth,clip]{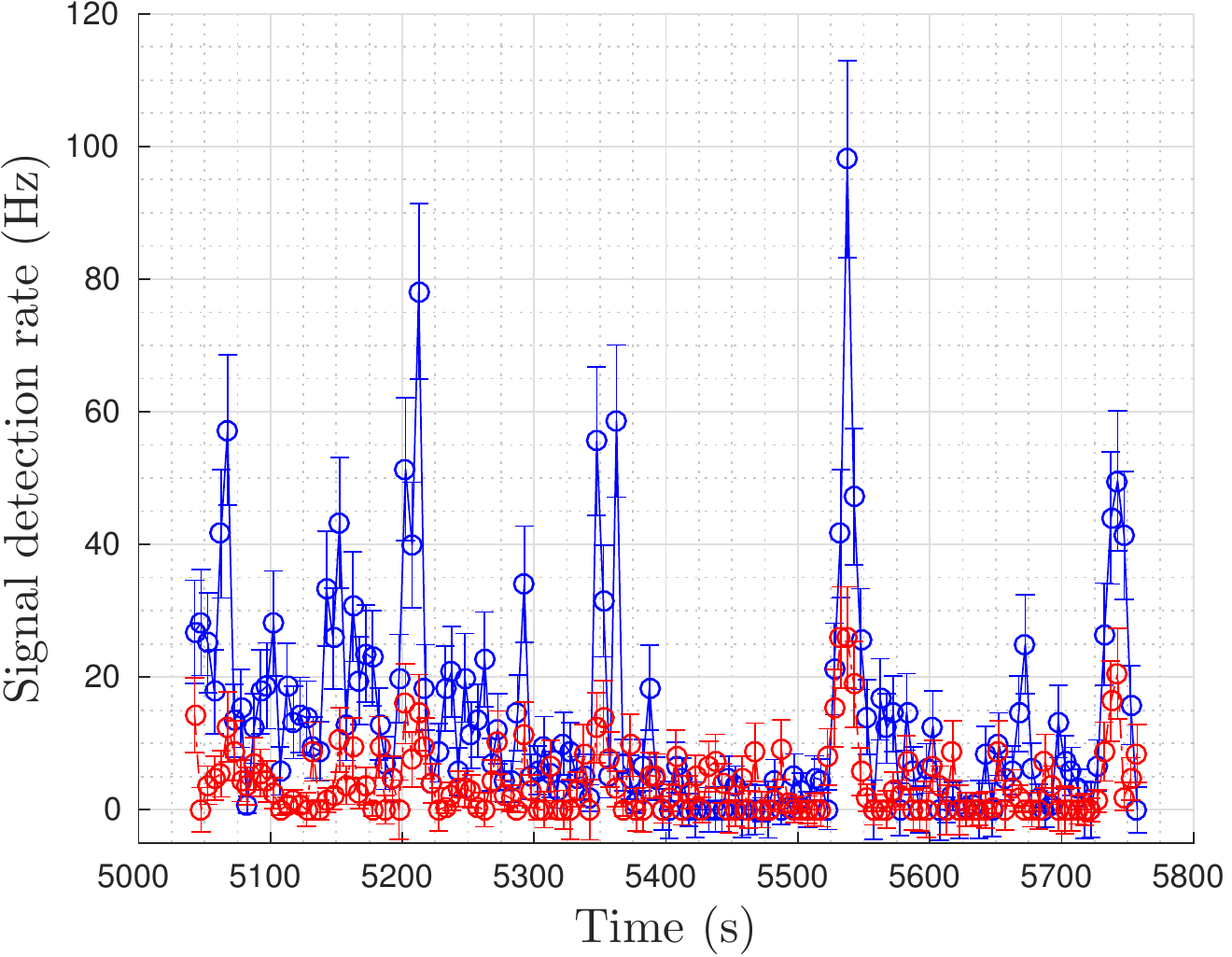}
    \caption{Signal detection rate for Glonass-131. The blue solid line refers to the MPD detector, while the red one refers to the PMT.}
    \label{fig:G131_det_rate}
\end{figure}

\textbf{Analysis of photons coming from Glonass-131.}
In the passage of Glonass-131, we used a slightly different receiver setup with respect to the one in Figure~\ref{fig:illustration}. Instead of using a single receiving detector, we placed a SPAD and a PMT detector (detection efficiency, $~$10$\%$; active diameter, 22 mm; H7360-02, Hamamatsu Photonics), coupled both to the down-going link with an additional 50:50 beam splitter. In this way we could compare the performances of the two detectors. Using the same analysis described in the main text, we obtained the signal detection rate presented in  Fig.~\ref{fig:G131_det_rate}. We noted a good correlation between the signal detection rate of the two detectors, although the PMT shows a much lower rate, since its quantum efficiency if five times lower than the one of SPAD. We then discarded the time windows with low signal detection rate, obtaining two comparable values for  $\overline{\mu}_{\mathrm{sat}}$ for the two detectors. The results are summarized in Tab.~\ref{tab:results}. We noted that MPD has about five times the signal rate of the PMT, as expected for the higher quantum efficiency. Also the SNR of the MPD is two times the SNR of the PMT, since the jitter on the time of arrival of the photons is lower.

\end{document}